\documentclass{article}
\newcommand{\bfr}{\begin{flushright}}
\newcommand{\efr}{\end{flushright}}
 
\begin{document}

\vspace*{1cm}

\begin{center}
{\bf \Large Color-kinematics duality and dimensional reduction for graviton emission in Regge limit}\footnote{Presented  by Agust{\' \i}n Sabio Vera at the Low x workshop, May 30 - June 4 2013, Rehovot and
Eilat, Israel}

\vspace*{0.5cm}
Henrik Johansson,$^{a,}$\footnote{\tt henrik.johansson@cern.ch}
Agust\'{\i}n Sabio Vera,$^{b,}$\footnote{\tt a.sabio.vera@gmail.com}
\\
Eduardo Serna Campillo,$^{c,}$\footnote{\tt eduardo.serna@usal.es} 
Miguel \' A. V\'azquez-Mozo$^{c, }$\footnote{\tt Miguel.Vazquez-Mozo@cern.ch}
\end{center}
\begin{center}
$^{a}${\sl Theory Division, Physics Department, CERN, 
CH-1211 Geneva 23, Switzerland
}

$^{b}${\sl Instituto de F\'{\i}sica Te\'orica UAM/CSIC \&
Universidad Aut\'onoma de Madrid\\
C/ Nicol\'as Cabrera 15, E-28049 Madrid, Spain
}

$^{c}${\sl Departamento de F\'{\i}sica Fundamental \& IUFFyM,
 Universidad de Salamanca \\ 
 Plaza de la Merced s/n,
 E-37008 Salamanca, Spain
  }
\end{center}

\date{\today
}
\begin{abstract}
In this talk we review the work in~\cite{SVSCVM,SVSCVM2,Johansson:2013nsa} where we have studied the 
applicability of the color-kinematics duality to the scattering of two distinguishable 
scalar matter particles with one gluon emission in QCD, or one graviton emission in Einstein gravity. We have shown that the duality works well in the Regge limit under two different extensions of the gauge theory: 
the introduction of a new scalar contact interaction and the relaxation of the 
distinguishability of the scalars.  Both modifications correspond to theories obtained by dimensional reduction from higher-dimensional pure gauge theories.
\end{abstract}

\section{Introduction}
The study of the relation between gravity and gauge theories both in the strong-weak aspect of AdS/CFT~\cite{AdS/CFT} and in a weak-weak set up~\cite{review_gauge_gravity} is an active field of research at present. 
Early results were the Kawai-Lewellen-Tye (KLT) relations~\cite{KLT} or the further studies in~\cite{BernGravity} which strengthened the idea that, at the level of scattering amplitudes, gravity should in some sense correspond to the ÒsquareÓ of gauge theory. More recently, Bern, Carrasco and one of the current authors (BCJ), showed that there exists an  underlying duality between color and kinematics in gauge theory~\cite{BCJ} which generates gravity amplitudes by replacing the color factors in the gauge-theory side with kinematic numerator functions depending on particle momenta and states, giving a double-copy representation of gravity amplitudes. This is expected to also work at  loop level~\cite{BCJLoop,BDHK,LoopNumerators,RecentLoopNumerators,N>=4SG,N=4SG}.  

The duality is known to work at tree level in pure (super)-Yang-Mills theories in various dimensions~\cite{BCJ,BDHK,ExplicitNumerators}. The treatment of general matter states and interactions in the color-kinematics duality has not been understood yet. 
In~\cite{SVSCVM2} three of us studied the duality in the context of inelastic amplitudes involving scalar particles in multi-Regge kinematics (\cite{Bartels:2008ce,Bartels:2008sc,Bartels:2012ra}). A simple extension of the BCJ duality to the scattering of two scalar particles with gluon emission in scalar QCD only correctly retrieves the square of two Lipatov's QCD emission vertices~\cite{BFKL1,BFKL2,BFKL3}. The terms responsible for the cancellation of simultaneous divergences in overlapping channels~\cite{Lipatov:2011ab,Lipatov:1982vv,Lipatov:1982it,Lipatov:1991nf,Bartels:2012ra,SVSCVM}, as required by unitarity~\cite{Steinmann} were not reproduced correctly. In~\cite{Johansson:2013nsa} this problem was approached with two different modifications: first, we considered the scattering of two distinguishable scalars in the adjoint representation in Yang-Mills theory introducing a quartic matter self-coupling (characteristic of the bosonic sector of $\mathcal{N}=2$ supersymmetric Yang-Mills theory). Second, we repeated the calculations in~\cite{SVSCVM2} with {\em identical} adjoint scalars. In both cases the duality reproduced the correct gravitational amplitude in the Regge limit computed in~\cite{SVSCVM}.


\section{Scalar matter and Color-kinematics duality 
\label{sec2}}
We focus on the scattering of two scalars with momenta $p_{1}$, $p_{2}$ producing two scalars with momentua $p_{3}$, $p_{4}$ and a emitted gluon (in QCD) or graviton (in gravity) with momentum $p_{5}$. The gauge-theory amplitude is written as a sum over 15 channels,
\begin{eqnarray}
\mathcal{A}_{5}=g^{3}\sum_{i=1}^{15}{c_{i}n_{i}\over d_{i}},
\label{eq:general_form}
\end{eqnarray}
where $c_{i}$ are the color factors:
\begin{eqnarray}
c_{1}={f}^{a_{5}a_{3}b}{f}^{ba_{4}c}{f}^{ca_{2}a_{1}}, & \hspace*{1cm} &
c_{2}={f}^{a_{5}a_{4}b}{f}^{ba_{3}c}{f}^{ca_{2}a_{1}}, \nonumber \\[0.2cm]
c_{3}={f}^{a_{2}a_{1}b}{f}^{ba_{5}c}{f}^{ca_{3}a_{4}}, & \hspace*{1cm} &
c_{4}={f}^{a_{5}a_{1}b}{f}^{ba_{2}c}{f}^{ca_{3}a_{4}}, \nonumber \\[0.2cm]
c_{5}={f}^{a_{5}a_{2}c}{f}^{ca_{1}b}{f}^{ba_{3}a_{4}}, & \hspace*{1cm} &
c_{6}={f}^{a_{5}a_{3}c}{f}^{ca_{1}b}{f}^{ba_{2}a_{4}}, \nonumber \\[0.2cm]
c_{7}={f}^{a_{5}a_{4}b}{f}^{ba_{2}c}{f}^{ca_{3}a_{1}}, & \hspace*{1cm} &
c_{8}={f}^{a_{5}a_{4}c}{f}^{ca_{1}b}{f}^{ba_{2}a_{3}}, \\[0.2cm]
c_{9}={f}^{a_{5}a_{3}b}{f}^{ba_{2}c}{f}^{ca_{4}a_{1}}, & \hspace*{1cm} &
c_{10}={f}^{a_{5}a_{1}b}{f}^{ba_{3}c}{f}^{ca_{2}a_{4}}, \nonumber \\[0.2cm]
c_{11}={f}^{a_{5}a_{2}b}{f}^{ba_{4}c}{f}^{ca_{3}a_{1}}, & \hspace*{1cm} &
c_{12}={f}^{a_{5}a_{2}b}{f}^{ba_{3}c}{f}^{ca_{4}a_{1}}, \nonumber \\[0.2cm]
c_{13}={f}^{a_{5}a_{1}b}{f}^{ba_{4}c}{f}^{ca_{2}a_{3}}, & \hspace*{1cm} &
c_{14}={f}^{a_{2}a_{4}b}{f}^{ba_{5}c}{f}^{ca_{3}a_{1}}, \nonumber \\[0.2cm]
c_{15}={f}^{a_{2}a_{3}b}{f}^{ba_{5}c}{f}^{ca_{4}a_{1}}, & \hspace*{1cm} &
\nonumber
\end{eqnarray}
where $f^{abc}$ are structure constants and the denominators $d_{i}=\prod_{\alpha_i}s_{\alpha_i}$ 
correspond to the product of the kinematic invariants associated with the internal lines in the corresponding 
Feynman diagram. The Jacobi identities of the structure constants make the color factors to satisfy nine independent identities that we label as $j_{\alpha}=0$ with
\begin{eqnarray}
j_{1} &\equiv& c_{12}-c_{9}+c_{15}, ~ j_{2} ~\equiv~ c_{11}-c_{7}+c_{14}, ~
j_{3} ~\equiv -c_{4}+c_{5}+c_{3}, \nonumber \\
j_{4} &\equiv& c_{1}-c_{2}-c_{3}, ~  j_{5}\equiv -c_{10}+c_{6}-c_{14}, ~ 
j_{6}\equiv -c_{13}+c_{8}-c_{15}, \label{eq:jacobi_identities}\\
j_{7} &\equiv& c_{4}-c_{10}+c_{13}, ~  j_{8}\equiv c_{8}+c_{7}-c_{2},~
j_{9}\equiv c_{6}+c_{9}-c_{1}.
\nonumber  
\end{eqnarray}
The numerators $n_{i}$ obtained from the Feynman rules in general do not satisfy the Jacobi-like identities
$\pm n_{i}\pm n_{j}\pm n_{k}=0$, corresponding to $j_{\alpha}$ with $c_i\rightarrow n_i$. A generalized gauge transformation, adding zero to the original amplitude in the form
\begin{eqnarray}
\mathcal{A}_{5}=\sum_{i=1}^{15}{c_{i}n_{i}\over d_{i}}+\sum_{\alpha=1}^{9}\gamma_{\alpha}j_{\alpha}
=\sum_{i=1}^{15}{c_{i}n'_{i}\over d_{i}}.
\end{eqnarray}
The numerators $n'_{i}$ are obtained by collecting 
the coefficients of each color factor $c_{i}$ and multiplying by corresponding denominator: $n'_{i}=d_i \partial_{c_i}{\cal A}_5$. The parameters $\gamma_{\alpha}$ are chosen such that $j_\alpha\Big|_{c_i\rightarrow n'_i} =0$.
These new numerators are used to construct the gravitational amplitude using the BCJ double-copy prescription
\begin{eqnarray}
-i\mathcal{M}=\left({\kappa\over 2}\right)^{3}\sum_{i=1}^{15}{n_{i}'\widetilde{n}'_{i}\over d_{i}},
\label{eq:gravamp_BCJ}
\end{eqnarray}
where $\kappa$ is the gravitational coupling constant. We express the momenta as
\begin{eqnarray}
p_{3}=-p_{1}+k_{1}, \hspace*{1cm}  p_{4}=-p_{2}-k_{2}, \hspace*{1cm} p_{5}=-k_{1}+k_{2},
\end{eqnarray}
where $k_{1}^{\mu}=\alpha_{1}p_{1}^{\mu}+\beta_{1}p_{2}^{\mu}+k_{1,\perp}^{\mu}$ and 
$k_{2}^{\mu}=\alpha_{2}p_{1}^{\mu}+\beta_{2}p_{2}^{\mu}+k_{2,\perp}^{\mu}$, with $k_{i,\perp}$ being orthogonal to $p_1$ and $p_2$. In this way we have
\begin{eqnarray}
p_{5}^{\mu}=(\alpha_{2}-\alpha_{1})p_{1}^{\mu}+(\beta_{2}-\beta_{1})p_{2}^{\mu}+k_{2,\perp}^{\mu}-k_{1,\perp}^{\mu}.
\end{eqnarray}
Multi-Regge kinematics is defined in terms of Sudakov parameters as $1\gg \alpha_{1} \gg \alpha_{2}$ and 
$1\gg |\beta_{2}|\gg |\beta_{1}|$. The gravitational amplitude then reads 
\begin{eqnarray}
-i\mathcal{M}=-iA_{kk}\mathcal{M}^{\mu\nu}\epsilon_{\mu\nu}(p_{5}),
\end{eqnarray}
where $\epsilon_{\mu\nu}(p_{5})$ is the graviton polarization tensor and \cite{SVSCVM}
\begin{eqnarray}
\mathcal{M}^{\mu\nu}&=&(k_{1}+k_{2})_{\perp}^{\mu}(k_{1}+k_{2})_{\perp}^{\nu}+
\mathcal{A}_{k1}\Big[(k_{1}+k_{2})_{\perp}^{\mu}p_{1}^{\nu}+p_{1}^{\mu}(k_{1}+k_{2})_{\perp}^{\nu}\Big] \nonumber \\
& & +\,\,\,\mathcal{A}_{k2}\Big[(k_{1}+k_{2})_{\perp}^{\mu}p_{2}^{\nu}+p_{2}^{\mu}(k_{1}+k_{2})_{\perp}^{\nu}\Big]
+\mathcal{A}_{12}\Big(p_{1}^{\mu}p_{2}^{\nu}+p_{2}^{\mu}p_{1}^{\nu}\Big) 
\label{eq:Mexpansion}
\\
& & +\,\,\,\mathcal{A}_{11}p_{1}^{\mu}p_{1}^{\nu}+\mathcal{A}_{22}p_{2}^{\mu}p_{2}^{\nu}.
\nonumber
\end{eqnarray}
The graviton emission effective vertex $\mathcal{M}^{\mu\nu}$ calculated by Lipatov in~\cite{Lipatov:1982vv} and by 
three of us more recently in~\cite{SVSCVM} is
\begin{eqnarray}
\mathcal{M}^{\mu\nu}=\Omega^{\mu}\Omega^{\nu}-\mathcal{N}^{\mu}\mathcal{N}^{\nu}.
\label{eq:lipatov_vertex}
\end{eqnarray}
where the vertex for the coupling of two reggeized gluons and one on-shell is 
\begin{eqnarray}
\Omega^{\mu}\simeq \left(\alpha_{1}-{2\beta_{1}\over \beta_{2}}\right)p^{\mu}
+\left(\beta_{2}+{2\alpha_{2}\over \alpha_{1}}\right)q^{\mu}-(k_{1}+k_{2})^{\mu}_{\perp},
\end{eqnarray}
and the term $\mathcal{N}^{\mu}\mathcal{N}^{\nu}$ removes simultaneous poles 
in $\alpha_{1}=0$ and $\beta_{2}=0$ with
\begin{eqnarray}
\mathcal{N}^{\mu}\simeq -2i\sqrt{\beta_{1}\alpha_{2}}\left({p^{\mu}\over\beta_{2}}+{q^{\mu}\over \alpha_{1}}
\right).
\end{eqnarray} 

In~\cite{SVSCVM2} three of us studied the scattering of two distinguishable scalars $\Phi$ and $\Phi'$ showing that the obtained color-kinematics solution only reproduces the QCD-like part ($\Omega^{\mu}\Omega^{\nu}$) in the gravity side. 
This problem with the incorrect $\mathcal{N}^{\mu}\mathcal{N}^{\nu}$ term was solved in~\cite{Johansson:2013nsa} by embedding the Yang-Mills + 2 scalar theory into the bosonic sector of 
$\mathcal{N}=2$ super-Yang-Mills theory (with the scalars transforming in the adjoint representation and introducing
the  matter self-coupling for the two scalars $\Delta\mathcal{L}={g^{2}\over 2}{\rm Tr\,}\Big([\Phi,\Phi']^{2}\Big)$). 
This adds four more diagrams without $t$-channel poles. Now the numerators are
\begin{eqnarray}
n_{1}'&=& (p_{1}+p_{2})^{2}\Big[-(\gamma_{9}-\gamma_{4})(p_{3}+p_{5})^{2}-2p_{3}\cdot\epsilon(p_{5})\Big],
\nonumber \\
n_{2}'&=& (p_{1}+p_{2})^{2}\Big[-(\gamma_{4}+\gamma_{8})(p_{4}+p_{5})^{2}+2p_{4}\cdot \epsilon(p_{5})\Big],
\nonumber \\
n_{3}'&=& (\gamma_{3}-\gamma_{4})(p_{1}+p_{2})^{2}(p_{3}+p_{4})^{2}, \nonumber  \\
n_{4}'&=& (p_{3}+p_{4})^{2}\Big[(\gamma_{7}-\gamma_{3})(p_{1}+p_{5})^{2}+2p_{1}\cdot \epsilon(p_{5})\Big], 
\nonumber \\
n_{5}'&=& -(p_{3}+p_{4})^{2}\Big[-\gamma_{3}(p_{2}+p_{5})^{2}+2p_{2}\cdot \epsilon(p_{5})\Big], \nonumber
\\
n_{6}'&=& -(p_{3}+p_{5})^{2}\Big[-(\gamma_{5}+\gamma_{9})(p_{2}+p_{4})^{2}+(p_{2}-p_{4})\cdot \epsilon(p_{5})\Big] 
\nonumber \\
& & -\,\,\,2(p_{2}-p_{4})\cdot (p_{1}-p_{3}-p_{5})[p_{3}\cdot\epsilon(p_{5})], 
\nonumber
\label{eq:numerators_disting}\\
n_{7}'&=& -(p_{4}+p_{5})^{2}\Big[(\gamma_{2}-\gamma_{8})(p_{1}+p_{3})^{2}+(p_{3}-p_{1})\cdot \epsilon(p_{5})\Big]
\nonumber \\
& &- \,\,\, 2(p_{3}-p_{1})\cdot (p_{2}-p_{4}-p_{5})[p_{3}\cdot\epsilon(p_{5})], 
\nonumber  \\
n_{8}'&=& (p_{2}+p_{3})^{2}\Big[(\gamma_{6}+\gamma_{8})(p_{4}+p_{5})^{2}+2p_{4}\cdot\epsilon(p_{5})\Big],
\label{eq:numdistprime} \\
n_{9}'&=& -(p_{1}+p_{4})^{2}\Big[(\gamma_{1}-\gamma_{9})(p_{3}+p_{5})^{2}+2p_{3}\cdot\epsilon(p_{5})\Big],
\nonumber \\
n_{10}'&=& -(p_{1}+p_{5})^{2}\Big[(\gamma_{5}+\gamma_{7})(p_{2}+p_{4})^{2}+(p_{2}-p_{4})\cdot\epsilon(p_{5})\Big]
\nonumber \\
& & -\,\,\, 2(p_{2}-p_{4})\cdot(-p_{1}+p_{3}-p_{5})[p_{1}\cdot\epsilon(p_{5})], \nonumber \\
n_{11}'&=& -(p_{2}+p_{5})^{2}\Big[-\gamma_{2}(p_{1}+p_{3})^{2}+(p_{3}-p_{1})\cdot\epsilon(p_{5})\Big] \nonumber \\
& & -\,\,\,2(p_{3}-p_{1})\cdot (-p_{2}+p_{4}-p_{5})[p_{2}\cdot\epsilon(p_{5})], \nonumber 
\\
n_{12}'&=& (p_{1}+p_{4})^{2}\Big[\gamma_{1}(p_{2}+p_{5})^{2}+2p_{2}\cdot\epsilon(p_{5})\Big], \nonumber \\
n_{13}'&=& -(p_{2}+p_{3})^{2}\Big[(\gamma_{6}-\gamma_{7})(p_{1}+p_{5})^{2}+2p_{1}\cdot\epsilon(p_{5})\Big], \nonumber 
\\
n_{14}'&=& -(p_{2}-p_{4})\cdot(p_{1}+p_{3}-p_{5})[(p_{3}-p_{1})\cdot \epsilon(p_{5})] \nonumber \\
& & -\,\,\,
(p_{3}-p_{1})\cdot (p_{2}-p_{4})[(-p_{1}+p_{2}-p_{3}+p_{4})\cdot\epsilon(p_{5})] \nonumber \\
& & +\,\,\, \gamma_{2}(p_{1}+p_{3})^{2}(p_{2}+p_{4})^{2}-\gamma_{5}(p_{1}+p_{3})^{2}(p_{2}+p_{4})^{2} ,\nonumber \\
n_{15}'&=& (\gamma_{1}-\gamma_{6})(p_{2}+p_{3})^{2}(p_{1}+p_{4})^{2}. \nonumber 
\end{eqnarray}
There are four independent $\gamma$ variables which we take to be $\gamma_{1,3,6,7}$ and write
\begin{eqnarray}
\gamma_{2} &=& {(p_{2}+2p_{3}+p_{4})\cdot\epsilon(p_{5})\over s\beta_{1}}
-\gamma_{1}{1+\beta_{2}\over \beta_{1}}-\gamma_{3}{-1+\alpha_{1}-\alpha_{2}+\beta_{1}-\beta_{2}\over \beta_{1}},
\nonumber \\[0.2cm]
\gamma_{4}&=& {2(p_{3}+p_{4})\cdot\epsilon(p_{5})\over s}+\gamma_{3}(1-\alpha_{1}+\alpha_{2}-\beta_{1}+\beta_{2})+
\gamma_{7}(\beta_{1}-\beta_{2}),  \\[0.2cm]
\gamma_{5}&=& {(-p_{2}+p_{4})\cdot\epsilon(p_{5})\over s\alpha_{2}}-\gamma_{3}
{1-\alpha_{1}+\alpha_{2}-\beta_{1}+\beta_{2}\over \alpha_{2}}+\gamma_{6}{1-\alpha_{1}\over \alpha_{2}}
-\gamma_{7}{\beta_{1}-\beta_{2}\over \alpha_{2}}, \nonumber \\[0.2cm]
\gamma_{8} &=& {2(p_{2}+p_{3})\cdot\epsilon(p_{5})\over s(\alpha_{1}+\beta_{1})}-\gamma_{1}{1+\beta_{2}\over
\alpha_{1}+\beta_{1}}+\gamma_{6}{1-\alpha_{1}\over \alpha_{1}+\beta_{1}}-\gamma_{7}{\beta_{1}-\beta_{2}\over
\alpha_{1}+\beta_{1}}, \nonumber \\[0.2cm]
\gamma_{9}&=& -{2(p_{2}+p_{3})\cdot\epsilon(p_{5})\over s(\alpha_{2}+\beta_{2})}+
\gamma_{1}{1+\beta_{2}\over \alpha_{2}+\beta_{2}}-\gamma_{6}{1-\alpha_{1}\over \alpha_{2}+\beta_{2}}.
\nonumber
\end{eqnarray}
Applying the BCJ prescription we construct the gravitational amplitude which in multi-Regge kinematics limit has the 
coefficients 
\begin{eqnarray}
\mathcal{A}_{11} &\simeq & \alpha_{1}^{2} -{4\alpha_{1}\beta_{1}\over \beta_{2}}
+{4\beta_{1}^{2}\over \beta_{2}^{2}}+{4\alpha_{2}\beta_{1}\over \beta_{2}^{2}}+\ldots\,,
\nonumber \\[0.2cm]
\mathcal{A}_{22} &\simeq & \beta_{2}^{2} +{4\alpha_{2}\beta_{1}\over \alpha_{1}}+{4\alpha_{2}\beta_{1}\over \alpha_{1}^{2}}
+{4\alpha_{2}^{2}\over \alpha_{1}^{2}}+\ldots\,,
\nonumber \\[0.2cm]
\mathcal{A}_{12} &\simeq & \alpha_{1}\beta_{2} -2\beta_{1}+2\alpha_{2}+\ldots\,,
\\[0.2cm]
\mathcal{A}_{k1} &\simeq & -\alpha_{1}+{2\beta_{1}\over \beta_{2}}+\ldots\,,
\nonumber \\[0.2cm]
\mathcal{A}_{k2} &\simeq & -\beta_{2}-{2\alpha_{2}\over \alpha_{1}}+\ldots\,,
\nonumber 
\label{coeffMRK}
\end{eqnarray}
which correctly reproduce the full form of Lipatov's effective graviton emission vertex.

A second method to avoid 
the problem found in~\cite{SVSCVM2} is to consider a single adjoint scalar minimally coupled to 
a nonabelian gauge field and compute the scattering amplitude with indistinguishable scalars.
The number of Feynman diagrams is increased and resolving the four-point vertices in terms of trivalent ones we obtain the numerators
\begin{eqnarray}
n_{1} &=& -
(p_{3}+p_{5})^{2}[(p_{2}-p_{1})\cdot\epsilon(p_{5})]
-2(p_{2}-p_{1})\cdot (-p_{3}+p_{4}-p_{5})[p_{3}\cdot\epsilon(p_{5})], \nonumber \\
n_{2}&=& -(p_{4}+p_{5})^{2}[(p_{2}-p_{1})\cdot\epsilon(p_{5})]
-2(p_{2}-p_{1})\cdot (p_{3}-p_{4}-p_{5})[p_{4}\cdot\epsilon(p_{5})], \nonumber \\
n_{3}&=& -(p_{2}-p_{1})\cdot(p_{3}-p_{4})[(p_{1}+p_{2}-p_{3}-p_{4})\cdot\epsilon(p_{5})] \nonumber \\
& & -\,\,\,(p_{3}-p_{4})\cdot(-p_{1}-p_{2}+p_{5})[(p_{2}-p_{1})\cdot\epsilon(p_{5})] \nonumber \\
& & -\,\,\,(p_{2}-p_{1})\cdot(p_{3}+p_{4}-p_{5})[(p_{3}-p_{4})\cdot\epsilon(p_{5})] ,\nonumber \\
n_{4}&=& -(p_{1}+p_{5})^{2}[(p_{3}-p_{4})\cdot \epsilon(p_{5})]-2(p_{3}-p_{4})\cdot(-p_{1}+p_{2}-p_{5})[p_{1}\cdot\epsilon(p_{5})], \nonumber \\
n_{5}&=& -(p_{2}+p_{5})^{2}[(p_{3}-p_{1})\cdot\epsilon(p_{5})]-2(p_{3}-p_{1})\cdot(-p_{2}+p_{4}-p_{5})
[p_{2}\cdot\epsilon(p_{5})], \nonumber \\
n_{6}&=& -(p_{3}+p_{5})^{2}[(p_{4}-p_{1})\cdot\epsilon(p_{5})]-2(p_{2}-p_{1})\cdot(-p_{3}+p_{4}-p_{5})
[p_{3}\cdot\epsilon(p_{5})], \nonumber  \\
n_{7}&=& -(p_{4}+p_{5})^{2}[(p_{3}-p_{1})\cdot \epsilon(p_{5})]-2(p_{3}-p_{1})\cdot(p_{2}-p_{4}-p_{5})
[p_{4}\cdot\epsilon(p_{5})],  \\
n_{8}&=& -(p_{4}+p_{5})^{2}[(p_{2}-p_{3})\cdot\epsilon(p_{5})]-2(p_{2}-p_{3})\cdot(p_{1}-p_{4}-p_{5})
[p_{4}\cdot \epsilon(p_{5})], \nonumber \\
n_{9}&=& -(p_{3}+p_{5})^{2}[(p_{4}-p_{1})\cdot\epsilon(p_{5})]-2(p_{4}-p_{1})\cdot(p_{2}-p_{3}-p_{5})
[p_{3}\cdot\epsilon(p_{5})], \nonumber \\
n_{10}&=& -(p_{1}+p_{5})^{2}[(p_{2}-p_{4})\cdot\epsilon(p_{5})]-2(p_{2}-p_{4})\cdot(-p_{1}+p_{3}-p_{5})
[p_{1}\cdot\epsilon(p_{5})], \nonumber \\
n_{11}&=& -(p_{2}+p_{5})^{2}[(p_{3}-p_{1})\cdot\epsilon(p_{5})]-2(p_{3}-p_{1})\cdot(-p_{2}+p_{4}-p_{5})
[p_{2}\cdot\epsilon(p_{5})], \nonumber \\
n_{12}&=& -(p_{2}+p_{5})^{2}[(p_{4}-p_{1})\cdot\epsilon(p_{5})]-2(p_{4}-p_{1})\cdot(-p_{2}+p_{3}-p_{5})
[p_{2}\cdot\epsilon(p_{5})], \nonumber \\
n_{13}&=& -(p_{1}+p_{5})^{2}[(p_{2}-p_{3})\cdot\epsilon(p_{5})]-2(p_{2}-p_{3})\cdot(-p_{1}+p_{4}-p_{5})
[p_{1}\cdot\epsilon(p_{5})], \nonumber \\
n_{14}&=& -(p_{2}-p_{4})\cdot(p_{1}+p_{3}-p_{5})[(p_{3}-p_{1})\cdot\epsilon(p_{5})] \nonumber \\
& & -\,\,\,(p_{3}-p_{1})\cdot(-p_{2}-p_{4}+p_{5})[(p_{2}-p_{4})\cdot\epsilon(p_{5})] \nonumber \\
& & -\,\,\,(p_{3}-p_{1})\cdot(p_{2}-p_{4})[(-p_{1}+p_{2}-p_{3}+p_{4})\cdot\epsilon(p_{5})], \nonumber\\
n_{15}&=& -(p_{2}-p_{3})\cdot(p_{4}-p_{1})[(-p_{1}+p_{2}+p_{3}-p_{4})\cdot\epsilon(p_{5})] \nonumber \\
& & -\,\,\,(p_{4}-p_{1})\cdot(-p_{2}-p_{3}+p_{5})[(p_{2}-p_{3})\cdot\epsilon(p_{5})] \nonumber \\
& & -\,\,\,(p_{2}-p_{3})\cdot(p_{1}+p_{4}-p_{5})[(p_{4}-p_{1})\cdot\epsilon(p_{5})].\nonumber
\end{eqnarray}
It is very interesting to point out that these numerators obtained from the Feynman rules directly satisfy the Jacobi-like identities and we do not have to perform a generalized gauge transformation before constructing the gravitational amplitude using the BCJ prescription. In the multi-Regge kinematics limit we find the same coefficients as in 
Eq.~(\ref{coeffMRK}) which again reproduce Lipatov's effective vertex.

\section{Discussion and outlook
\label{sec5}}
In this contribution we have summarized the work in~\cite{SVSCVM,SVSCVM2,Johansson:2013nsa} related to the use of the  color-kinematics duality to the 
scattering of two distinguishable scalar matter particles with gluon emission, or graviton emission. In~\cite{SVSCVM2}
it was shown that in transferring the BCJ double-copy prescription to the scattering of minimally coupled distinguishable scalars an important part of the gravitational amplitude in multi-Regge kinematics was not correctly reproduced.

In~\cite{Johansson:2013nsa} we have studied two extensions of the theory for which the BCJ prescription generates the correct the Regge limit of~\cite{Lipatov:1982vv,SVSCVM}. In one of them a contact interaction between the two scalar particles is introduced, while in the other we give up the distinguishability of the scalars. For both cases we obtain  valid gravity amplitudes from the BCJ double-copy prescription in the Regge limit. 
  
Both cases can be thought of as originating from the bosonic sector of $D=4$ $\mathcal{N}=2$ super-Yang-Mills theory, keeping either both scalars, or only one scalar. They can be interpreted as coming from subsectors of $\mathcal{N}=4$ super-Yang-Mills theory, for which double-copy prescription is proven to give valid gravity tree-level amplitudes~\cite{BDHK}. An important observation is that the $D = 4$ Yang-Mills + scalar theories studied in~\cite{Johansson:2013nsa} are via dimensional reduction directly related to pure Yang-Mills theory in $D=6$ and $D=5$ dimensions, respectively. 
Indeed the new interaction term can be obtained by dimensionally reducing $D=6$ pure Yang-Mills to $D=4$, 
where the gauge field along the extra two dimensions are interpreted as two scalars,
$\Phi\equiv A_{4}$, $\Phi'\equiv A_{5}$. We find that the successful application of color-kinematics duality in~\cite{Johansson:2013nsa} stems from its validity in higher-dimensional Yang-Mills theory and gravity~\cite{BCJ,BCJLoop,BDHK,LoopNumerators,N=4SG} (other practical examples of dimensional reduction can be found in~\cite{vm_SYM}).

Nevertheless, the inclusion of general matter states and interactions in the color-kinematics and double-copy formalism is still an open problem. In particular, it would be interesting to study how to relate tree amplitudes in Yang-Mills theory with minimally-coupled fermions and scalars to that of Einstein gravity with the same matter content. Embedding the gauge and gravity theories into their respective higher-dimensional versions is probably the correct path to follow. The results here discussed with the help of the multi-Regge limit are a first step towards understanding this general matter case.

\section*{Acknowledgments} 

A.S.V. acknowledges partial support from the European Comission under contract LHCPhenoNet (PITN-GA-2010-264564), 
the Madrid Regional Government through Proyecto HEPHACOS ESP-1473, the Spanish Government 
MICINN (FPA2010-17747) and Spanish MINECOs ``Centro de Excelencia Severo Ochoa" Programme under grant  SEV-2012-0249. 
The work of E.S.C. 
has been supported by a Spanish Government FPI Predoctoral Fellowship and grant FIS2012-30926. 
M.A.V.-M. acknowledges partial support from Spanish Government grants FPA2012-34456 and FIS2012-30926, 
Basque Government Grant IT-357-07 and Spanish Consolider-Ingenio 2010 Programme CPAN (CSD2007-00042).
M.A.V.-M. thanks the Instituto de F\'{\i}sica Te\'orica UAM/CSIC for kind hospitality. 


\bibliographystyle{apsrev4-1}

\begin{thebibliography}{99}


\bibitem{SVSCVM}
A.~Sabio Vera, E.~Serna~Campillo and M.~A.~V\'azquez-Mozo,
  {\it Graviton emission in Einstein-Hilbert gravity,}
  J. High Energy Phys {\bf 03} (2012) 005
 {\tt [arXiv:1112.4494 [hep-th]].}

 \bibitem{SVSCVM2}
A.~Sabio~Vera, E.~Serna~Campillo and M.~A.~V\'azquez-Mozo,
  {\it Color-Kinematics Duality and the Regge Limit of Inelastic Amplitudes,}
  J.\  High Energy Phys.\  {\bf 04} (2013) 086
{\tt  [arXiv:1212.5103 [hep-th]]}.
  
\bibitem{Johansson:2013nsa}
  H.~Johansson, A.~Sabio~Vera, E.~Serna~Campillo and M.~A.~V\'azquez-Mozo,
  {\it Color-Kinematics Duality in Multi-Regge Kinematics and Dimensional Reduction,}
  arXiv:1307.3106 [hep-th], to be published in JHEP.
  
    \bibitem{AdS/CFT}
J.~M.~Maldacena,
  {\it The Large N limit of superconformal field theories and supergravity,}
  Adv.\ Theor.\ Math.\ Phys.\  {\bf 2} (1998) 231
  {\tt [hep-th/9711200]};\\
  %
  S.~S.~Gubser, I.~R.~Klebanov and A.~M.~Polyakov,
  {\it Gauge theory correlators from noncritical string theory,}
  Phys.\ Lett.\ {\bf B428} (1998) 105
  {\tt [hep-th/9802109]};\\
  %
  E.~Witten,
  {\it Anti-de Sitter space and holography,}
  Adv.\ Theor.\ Math.\ Phys.\  {\bf 2} (1998) 253
  {\tt [hep-th/9802150]};\\
  %
  O.~Aharony, S.~S.~Gubser, J.~M.~Maldacena, H.~Ooguri and Y.~Oz,
  {\it Large N field theories, string theory and gravity,}
  Phys.\ Rept.\  {\bf 323} (2000) 183
 {\tt [hep-th/9905111].}

  \bibitem{review_gauge_gravity}
 Z.~Bern,
  {\it Perturbative quantum gravity and its relation to gauge theory,}
  Living Rev.\ Rel.\  {\bf 5} (2002) 5
  {\tt [gr-qc/0206071].}

\bibitem{KLT}
H.~Kawai, D.~C.~Lewellen and S.~H.~H.~Tye,
  {\it A Relation Between Tree Amplitudes of Closed and Open Strings,}
  Nucl.\ Phys.\ {\bf B269} (1986) 1.
%

\bibitem{BernGravity}
  Z.~Bern, D.~C.~Dunbar,
  {\it A Mapping between Feynman and string motivated one loop rules in gauge
  theories,}
  Nucl.\ Phys.\  {\bf B379} (1992) 562;\\
 %
  Z.~Bern, D.~C.~Dunbar and T.~Shimada,
  {\it String based methods in perturbative gravity,}
  Phys.\ Lett.\  {\bf B312} (1993) 277
{\tt [arXiv:hep-th/9307001]};\\
 %
  D.~C.~Dunbar and P.~S.~Norridge,
  {\it Calculation of graviton scattering amplitudes using string based methods,}
  Nucl.\ Phys.\  {\bf B433} (1995) 181
{\tt  [arXiv:hep-th/9408014]};\\
  %
  Z.~Bern, A.~De Freitas and H.~L.~Wong,
  {\it On the coupling of gravitons to matter,}
  Phys.\ Rev.\ Lett.\  {\bf 84} (2000) 3531
  {\tt [hep-th/9912033]};\\
  %
  Z.~Bern and A.~K.~Grant,
  {\it Perturbative gravity from QCD amplitudes,}
  Phys.\ Lett.\ B {\bf 457} (1999) 23
  {\tt [hep-th/9904026]}.

  \bibitem{BCJ}
Z.~Bern, J.~J.~M.~Carrasco and H.~Johansson,
  {\it New Relations for Gauge-Theory Amplitudes,}
  Phys.\ Rev.\ {\bf D78} (2008) 085011
  {\tt [arXiv:0805.3993 [hep-ph]]}.

\bibitem{BCJLoop}
  Z.~Bern, J.~J.~M.~Carrasco and H.~Johansson,
  {\it Perturbative Quantum Gravity as a Double Copy of Gauge Theory,}
  Phys.\ Rev.\ Lett.\  {\bf 105} (2010) 061602.
{\tt [arXiv:1004.0476 [hep-th]].}

  \bibitem{BDHK}
Z.~Bern, T.~Dennen, Y.-t.~Huang and M.~Kiermaier,
  {\it Gravity as the Square of Gauge Theory,}
  Phys.\ Rev.\ {\bf D82} (2010) 065003
  {\tt [arXiv:1004.0693 [hep-th]].}
  
  \bibitem{ExplicitNumerators}
M. Kiermaier, presented at Amplitudes 2010,\\
http://www.strings.ph.qmul.ac.uk/\~{}theory/Amplitudes2010/;\\
%
N.~E.~J.~Bjerrum-Bohr, P.~H.~Damgaard, T.~Sondergaard and P.~Vanhove,
JHEP {\bf 1101}, 001 (2011)
[arXiv:1010.3933 [hep-th]];\\
%
C.~R.~Mafra, O.~Schlotterer and S.~Stieberger,
JHEP {\bf 1107}, 092 (2011)
[arXiv:1104.5224 [hep-th]].



  \bibitem{LoopNumerators}
    J.~J.~Carrasco and H.~Johansson,
  {\it Five-Point Amplitudes in N=4 Super-Yang-Mills Theory and N=8 Supergravity,}
  Phys.\ Rev.\ {\bf D85} (2012) 025006
  {\tt [arXiv:1106.4711 [hep-th]]};\\
  %
  Z.~Bern, J.~J.~M.~Carrasco, L.~J.~Dixon, H.~Johansson and R.~Roiban,
  {\it Simplifying Multiloop Integrands and Ultraviolet Divergences of Gauge Theory and Gravity Amplitudes,}
  Phys.\ Rev.\ {\bf D85} (2012) 105014
 {\tt [arXiv:1201.5366 [hep-th]].}


 \bibitem{RecentLoopNumerators} 
  %
   R.~H.~Boels, B.~A.~Kniehl, O.~V.~Tarasov and G.~Yang,
  {\it Color-kinematic Duality for Form Factors,}
  J. High Energy Phys. {\bf 02} (2013) 063
 {\tt  [arXiv:1211.7028 [hep-th]]};\\
  %
  J.~J.~M.~Carrasco, M.~Chiodaroli, M.~Gunaydin and R.~Roiban,
  {\it One-loop four-point amplitudes in pure and matter-coupled $N \leq 4$ supergravity,}
  J. High Energy Phys. {\bf 03} (2013) 056
  {\tt [arXiv:1212.1146 [hep-th]]};\\
  %
  R.~H.~Boels, R.~S.~Isermann, R.~Monteiro and D.~O'Connell,
  {\it Colour-Kinematics Duality for One-Loop Rational Amplitudes,}
  J. High Energy Phys {\bf 04} (2013) 107
  {\tt [arXiv:1301.4165 [hep-th]]};\\
  %
    N.~E.~J.~Bjerrum-Bohr, T.~Dennen, R.~Monteiro and D.~O'Connell,
  {\it Integrand Oxidation and One-Loop Colour-Dual Numerators in $N=4$ Gauge Theory,}
 {\tt arXiv:1303.2913 [hep-th]};\\
  %
  Z.~Bern, S.~Davies, T.~Dennen, Y.~-t.~Huang and J.~Nohle,
  {\it Color-Kinematics Duality for Pure Yang-Mills and Gravity at One and Two Loops,}
 {\tt arXiv:1303.6605 [hep-th]}\,.
 
   \bibitem{N>=4SG} 
  Z.~Bern, C.~Boucher-Veronneau and H.~Johansson,
  {\it $N \geq 4$ Supergravity Amplitudes from Gauge Theory at One Loop,}
  Phys.\ Rev.\ {\bf D84} (2011) 105035
 {\tt [arXiv:1107.1935 [hep-th]]};\\
  %
  C.~Boucher-Veronneau and L.~J.~Dixon,
  {\it $N \geq 4$ Supergravity Amplitudes from Gauge Theory at Two Loops,}
  J. High Energy Phys. {\bf 12} (2011) 046
  {\tt [arXiv:1110.1132 [hep-th]].}
 
  
\bibitem{N=4SG} 
Z.~Bern, S.~Davies, T.~Dennen and Y.~-t.~Huang,
  {\it Absence of Three-Loop Four-Point Divergences in $N=4$ Supergravity,}
  Phys.\ Rev.\ Lett.\  {\bf 108} (2012) 201301
  {\tt [arXiv:1202.3423 [hep-th]]};\\
  %
  Z.~Bern, S.~Davies, T.~Dennen and Y.~-t.~Huang,
  {\it Ultraviolet Cancellations in Half-Maximal Supergravity as a Consequence of the Double-Copy Structure,}
  Phys.\ Rev.\ {\bf D86} (2012) 105014
 {\tt [arXiv:1209.2472 [hep-th]]};\\
  %
 Z.~Bern, S.~Davies and T.~Dennen,
  {\it The Ultraviolet Structure of Half-Maximal Supergravity with Matter Multiplets at Two and Three Loops,}
 {\tt arXiv:1305.4876 [hep-th].}
 



\bibitem{Bartels:2008ce}
  J.~Bartels, L.~N.~Lipatov and A.~Sabio Vera,
  {\it BFKL Pomeron, Reggeized gluons and Bern-Dixon-Smirnov amplitudes,}
  Phys.\ Rev.\  {\bf D80} (2009) 045002
{\tt  [arXiv:0802.2065 [hep-th]].}

\bibitem{Bartels:2008sc}
  J.~Bartels, L.~N.~Lipatov and A.~Sabio Vera,
  {\it $\mathcal{N}=4$ supersymmetric Yang Mills scattering amplitudes at high energies: The
  Regge cut contribution,}
  Eur.\ Phys.\ J.\  {\bf C65} (2010) 587
{\tt [arXiv:0807.0894 [hep-th]].}
  

\bibitem{Bartels:2012ra}
  J.~Bartels, L.~N.~Lipatov and A.~Sabio~Vera,
  {\it Double-logarithms in Einstein-Hilbert gravity and supergravity,}
{\tt  arXiv:1208.3423 [hep-th].}

  \bibitem{BFKL1}  
L.~N.~Lipatov, 
{\it  	
Reggeization of the Vector Meson and the Vacuum Singularity in Nonabelian Gauge Theories},
Sov.\ J.\ Nucl.\ Phys.\  {\bf 23} (1976) 338.

\bibitem{BFKL2}
E.~A.~Kuraev, L.~N.~Lipatov and V.~S.~Fadin, {\it 	
On the Pomeranchuk Singularity in Asymptotically Free Theories},
Phys.\ Lett.\  B {\bf 60} (1975) 50, 
\\
E.~A.~Kuraev, L.~N.~Lipatov and V.~S.~Fadin,
{\it  	
Multi-Reggeon Processes in the Yang-Mills Theory},
Sov.\ Phys.\ JETP {\bf 44} (1976) 443,
\\
E.~A.~Kuraev, L.~N.~Lipatov and V.~S.~Fadin,
{\it The Pomeranchuk Singularity in Nonabelian Gauge Theories},
Sov.\ Phys.\ JETP {\bf 45} (1977) 199.

\bibitem{BFKL3}
I.~I.~Balitsky and L.~N.~Lipatov,
{\it  	
The Pomeranchuk Singularity in Quantum Chromodynamics}, 
Sov.\ J.\ Nucl.\ Phys.\  {\bf 28} (1978) 822. 


\bibitem{Lipatov:2011ab}
  L.~N.~Lipatov,
  {\it Effective action for the Regge processes in gravity,}
  Phys.\ Part.\ Nucl.\  {\bf 44} (2013) 391
  {\tt [arXiv:1105.3127 [hep-th]].}

\bibitem{Lipatov:1982vv}
  L.~N.~Lipatov,
  {\it Graviton Reggeization,}
  Phys.\ Lett.\ {\bf B116} (1982) 411.
  

\bibitem{Lipatov:1982it}
  L.~N.~Lipatov,
  {\it Multi-Regge Processes In Gravitation,}
  Sov.\ Phys.\ JETP {\bf 55} (1982) 582
   [Zh.\ Eksp.\ Teor.\ Fiz.\  {\bf 82} (1982) 991].
 

\bibitem{Lipatov:1991nf}
  L.~N.~Lipatov,
  {\it High-energy scattering in QCD and in quantum gravity and two-dimensional field theories,}
  Nucl.\ Phys.\ {\bf B365} (1991) 614.
  
  

  
  
  
\bibitem{Steinmann}
O.~Steinmann, 
{\it \"Uber den Zusammenhang zwischen den Wightmanfunktionen 
und der retardierten Kommutatoren}, 
Helv. Phys. Acta {\bf 33} (1960) 257, 
\\
O.~Steinmann, {\it Wightman-Funktionen und retardierten Kommutatoren. II}, 
Helv. Phys. Acta. {\bf 33} (1960) 347.



\bibitem{vm_SYM}
M.~A.~V\'azquez-Mozo,
  {\it A Note on supersymmetric Yang-Mills thermodynamics,}
  Phys.\ Rev.\ {\bf D60} (1999) 106010
 {\tt [hep-th/9905030].}
  
  




\end{thebibliography}


\end{document}